\documentclass[11pt]{article} 
\usepackage{amsfonts, amssymb, amsmath, amsthm, graphicx, hyperref, caption, xcolor, verbatim, titlesec, booktabs, multirow, lscape, enumitem, setspace, tocloft, natbib, subfig}
\usepackage{appendix}
\usepackage[sans]{dsfont}
\usepackage[top=0.9in,bottom=0.9in,left=0.9in,right=0.9in]{geometry} 
\hypersetup{pdfborder = {0 0 0},colorlinks=true,linkcolor=blue,urlcolor=blue,citecolor=blue}
\usepackage[ruled,vlined]{algorithm2e}
\newtheoremstyle{theoremstyle}
  {\topsep}{\topsep}{\itshape}{}{}{}{.5em}
  {\color{blue}\ifthenelse{\equal{#3}{}}{{\bfseries #1 #2.}}{{\bfseries #1 #2. (#3)}}}
\newtheoremstyle{examplestyle}
  {\topsep}{\topsep}{}{}{}{}{.5em}
  {\color{blue}\ifthenelse{\equal{#3}{}}{{\bfseries #1 #2.}}{{\bfseries #1 #2. (#3)}}}
\theoremstyle{theoremstyle}\newtheorem{thm}{Theorem}
\theoremstyle{theoremstyle}     
\theoremstyle{theoremstyle}\newtheorem{lem}{Lemma}  
\theoremstyle{theoremstyle}        
\theoremstyle{theoremstyle}\newtheorem{prop}{Proposition}
\theoremstyle{theoremstyle}\newtheorem{assumption}{Assumption}

\theoremstyle{examplestyle}
\theoremstyle{examplestyle}\newtheorem{remark}{Remark}
\theoremstyle{examplestyle}\newtheorem{algo}{Algorithm}

\newcommand{\EndOfTheorem}{\qquad $\parallel$}
\newcommand{\Expectation}{\mathbb{E}}

\newcommand{\Var}{\mathbb{V}}

\newcommand{\Prob}{\mathbb{P}}
\newcommand{\Indicator}{\mathds{1}}
\newcommand{\op}{o_{\mathrm{p}}}
\newcommand{\Op}{O_{\mathrm{p}}}

\newcommand{\toProb}{\overset{\mathrm{p}}{\to}}
\newcommand{\toDist}{\overset{\mathrm{d}}{\to}}

\newcommand{\Trans}{\mathrm{T}}
\newcommand{\diff}{\mathrm{d}}

\renewcommand{\epsilon}{\varepsilon}

\DeclareMathOperator*{\argmin}{argmin}

\newcommand{\Bias}{\mathsf{B}}
\newcommand{\Variance}{\mathsf{V}}

\begin{document}

\title{ 
Robust Inference Using Inverse Probability Weighting\thanks{
The authors are deeply grateful to Matias Cattaneo for the comments and suggestions that significantly improved the manuscript. The authors also thank Sebastian Calonico, Max Farrell, Yingjie Feng, Andreas Hagemann, Xuming He, Michael Jansson, Lutz Kilian, Jose Luis Montiel Olea, Kenichi Nagasawa, Roc\'{i}o Titiunik, Gonzalo Vazquez-Bare, the editor, an associate editor, and two referees for their valuable feedback and thoughtful discussions.}
}
\author{
Xinwei Ma\thanks{Department of Economics, University of California, San Diego.} 
\and
Jingshen Wang\thanks{Division of Biostatistics, University of California, Berkeley.} 
}
\maketitle

\begin{abstract}

\noindent Inverse Probability Weighting (IPW) is widely used in empirical work in economics and other disciplines. As Gaussian approximations perform poorly in the presence of ``small denominators,'' trimming is routinely employed as a regularization strategy. However, ad hoc trimming of the observations renders usual inference procedures invalid for the target estimand, even in large samples. In this paper, we first show that the IPW estimator can have different (Gaussian or non-Gaussian) asymptotic distributions, depending on how ``close to zero'' the probability weights are and on how large the trimming threshold is. As a remedy, we propose an inference procedure that is robust not only to small probability weights entering the IPW estimator but also to a wide range of trimming threshold choices, by adapting to these different asymptotic distributions. This robustness is achieved by employing resampling techniques and by correcting a non-negligible trimming bias. We also propose an easy-to-implement method for choosing the trimming threshold by minimizing an empirical analogue of the asymptotic mean squared error. In addition, we show that our inference procedure remains valid with the use of a data-driven trimming threshold. We illustrate our method by revisiting a dataset from the National Supported Work program.

\vskip 1.5em

\end{abstract}

\vfill

\thispagestyle{empty}
\clearpage

\doublespacing
\setcounter{page}{1}
\pagestyle{plain}

\section{Introduction}\label{section-1:introduction}
Inverse Probability Weighting (IPW) is widely used in empirical work in economics and other disciplines. In practice, it is common to observe small probability weights entering the IPW estimator. This renders inference based on standard Gaussian approximations invalid, even in large samples, because these approximations rely crucially on the probability weights being well-separated from zero. In a recent study, \cite*{busso2014finite} investigated the finite sample performance of commonly used IPW treatment effect estimators, and documented that small probability weights can be detrimental to statistical inference. In response to this problem, observations with probability weights below a certain threshold are often excluded from subsequent statistical analysis. The exact amount of trimming, however, is usually ad hoc and will affect the performance of the IPW estimator and the corresponding confidence interval in nontrivial ways.

In this paper, we show that the IPW estimator can have different (Gaussian or non-Gaussian) asymptotic distributions, depending on how ``close to zero'' the probability weights are and on how large the trimming threshold is. We propose an inference procedure that adapts to these different asymptotic distributions, making it robust not only to small probability weights, but also to a wide range of trimming threshold choices. This ``two-way robustness'' is achieved by combining subsampling with a novel bias correction technique. In addition, we propose an easy-to-implement method for choosing the trimming threshold by minimizing an empirical analogue of the asymptotic mean squared error, and show that our inference procedure remains valid with the use of a data-driven trimming threshold. 

To understand why standard inference procedures are not robust to small probability weights, and why their performance can be sensitive to the amount of trimming, we first study the large-sample properties of the IPW estimator
\begin{align*}
\hat{\theta}_{n,b_n} = \frac{1}{n}\sum_{i=1}^n \frac{D_iY_i}{\hat{e}( X_i)}\Indicator_{\hat{e}(X_i)\geq b_n},
\end{align*}
where $D_i\in\{0,1\}$ is binary, $Y_i$ is the outcome of interest, $e( X_i)=\Prob[D_i=1| X_i]$ is the probability weight conditional on the covariates with $\hat{e}(X_i)$ being its estimate, and $b_n$ is the trimming threshold (the untrimmed IPW estimator is a special case with $b_n=0$). The asymptotic framework we employ is general and allows, but does not require that the probability weights have a heavy tail near zero. Specifically, if the tail is relatively thin, the asymptotic distribution will be Gaussian; otherwise a slower-than-$\sqrt{n}$ convergence rate and a non-Gaussian asymptotic distribution can emerge, and they will depend on the trimming threshold $b_n$. In the latter case,
\begin{align}\label{eq:trimmed IPW limiting distribution}
\frac{n}{a_{n,b_n}}\Big(\hat\theta_{n,b_n} - \theta_0 - \Bias_{n,b_n} \Big) \toDist \mathcal{L}(\gamma_0,\alpha_+(\cdot),\alpha_-(\cdot)),
\end{align} 
where $\theta_0$ is the parameter of interest, $a_{n,b_n}\to \infty$ is a sequence of normalizing factors, and $\toDist$ denotes convergence in distribution. 

First, a trimming bias $\Bias_{n,b_n}$ emerges. This bias has order $\Prob[e(X)\leq b_n]$, and hence it will vanish asymptotically if the trimming threshold shrinks to zero. What matters for inference, however, is the asymptotic bias, defined as the trimming bias scaled by the convergence rate: $\frac{n}{a_{n,b_n}}\mathsf{B}_{n,b_n}$. This asymptotic bias may not vanish even in large samples, and can be detrimental to statistical inference, as it shifts the asymptotic distribution away from the target estimand. Second, the asymptotic distribution, $\mathcal{L}(\cdot)$, depends on three parameters. The first parameter $\gamma_0$ is related to tail behaviors of the probability weights near zero, and the other two parameters characterize shape and tail properties of the asymptotic distribution. In particular, $\mathcal{L}(\cdot)$ does not need to be symmetric. Third, the convergence rate, $n/a_{n,b_n}$, is usually unknown, and depends again on how ``close to zero'' the probability weights are and how large the trimming threshold is.  

As the large-sample properties of the IPW estimator are sensitive to small probability weights and to the amount of trimming, it is important to develop an inference procedure that automatically adapts to the relevant asymptotic distributions. However, the presence of additional nuisance parameters makes it challenging to base inference on estimating the asymptotic distribution. In addition, the standard nonparametric bootstrap is known to fail in our setting \citep{athreya1987bootstrap, knight1989bootstrap}. We instead propose the use of subsampling \citep{politis1994large}. We show that subsampling provides valid approximations to the asymptotic distribution in \eqref{eq:trimmed IPW limiting distribution}. With self-normalization (i.e., subsampling a Studentized statistic), it also overcomes the difficulty of having a possibly unknown convergence rate. Subsampling alone does not suffice for valid inference due to the bias induced by trimming. To make our inference procedure also robust to a wide range of trimming threshold choices, we combine subsampling with a novel bias correction method based on local polynomial regressions \citep{fan-Gijbels_1996_Book}. To be precise, our method regresses the outcome variable on a polynomial basis of the probability weight in a region local to the origin, and estimates the trimming bias with the regression coefficients. 

We also address the question of how to choose the trimming threshold. One extreme possibility is fixed trimming ($b_n=b>0$). Although fixed trimming helps restore asymptotic Gaussianity by forcing the probability weights to be bounded away from zero, this practice is difficult to justify, unless one is willing to re-interpret the estimation and inference result completely. We instead propose to determine the trimming threshold by taking into account the bias and variance of the (trimmed) IPW estimator. We suggest an easy-to-implement method to choose the trimming threshold by minimizing an empirical analogue of the asymptotic mean squared error. 

This paper relates to a large body of literature on program evaluation and causal inference \citep{imbens2015causal, abadie2018econometric, hernan2018book}. Estimators with inverse weighting are widely used in missing data models \citep*{robins1994estimation, wooldridge2007inverse} and treatment effect estimation \citep*{hirano2003efficient, cattaneo2010efficient}. They also feature in settings such as instrumental variables \citep{abadie2003semiparametric}, difference-in-differences \citep{abadie2005semiparametric}, and counterfactual analysis \citep*{dinardo1996labor}. \cite{khan2010irregular} show that, depending on tail behaviors of the probability weights, the variance bound of the IPW estimator can be infinite, which leads to a slower-than-$\sqrt{n}$ convergence rate.  \cite{sasaki2018ratio} propose a trimming method and a companion sieve-based bias correction technique for conducting inference for moments of ratios, which complement our paper. \cite{chaudhuri2014heavy} propose a different trimming strategy based on $|DY/e( X)|$ rather than the probability weight, and an inference procedure relying on asymptotic Gaussianity. \cite*{crump2009dealing} also study the problem of trimming threshold selection, the difference is that their method is based on minimizing a variance term, and hence can lead to a much larger trimming threshold than what we propose. 

The untrimmed IPW estimator (i.e., $b_n=0$) is a special case of \eqref{eq:trimmed IPW limiting distribution}, and the asymptotic distribution is known as the L\'{e}vy stable distribution. Stable convergence has been established in many contexts. For example,  \cite{vaynman2014stable} show that stable convergence may arise for the variance targeting estimator, and hence the tail trimming of \cite{hill2012variance} can be crucial for establishing asymptotic Gaussianity. \cite{khan2013Uniform} also establish a stable convergence result for the untrimmed IPW estimator. However, they do not discuss the impact of trimming or how the trimming threshold should be chosen in practice. \cite*{hong2018inference} consider a different setting where observations fall into finitely many strata. They demonstrate that for estimating treatment effects the effective sample size can be much smaller as a result of disproportionately many treated or control units (a.k.a. limited overlap), and relate the rate of convergence to how fast the probability weight approaches an extreme. 

With the IPW estimator as a special case, \cite{cattaneo2018kernelBased} and \cite*{cattaneo2018manyCovs} show how an asymptotic bias can arise in a two-step semiparametric setting as a result of overfitting the first step. \cite*{chernozhukov2018locally} develop robust inference procedures against underfitting bias. The trimming bias we document in this paper is both qualitatively and quantitatively different, as it will be present even when the probability weights are observed, and certainly will not disappear with model selection or machine learning methods \citep*{athey2018approximate,belloni2018HighDimensional, farrell2015Robust,farrell2018neural}.

The rest of the paper is structured as follows. In Section \ref{section-2: large sample properties}, we state and discuss the main assumptions, and study the large-sample properties of the IPW estimator. In Section \ref{section-3: robust inference}, we discuss in detail our robust inference procedure, including how the bias correction and the subsampling are implemented. A data-driven method to choose the trimming threshold is also proposed. Section \ref{section-4:empirical} showcases our methods with an empirical example. Section \ref{section:conclusion} concludes. To conserve space, we collect auxiliary lemmas, additional results, simulation evidence, and all proofs in the online Supplementary Material. We also discuss in the Supplementary Material how our IPW framework can be generalized to provide robust inference for treatment effect estimands and parameters defined by general (nonlinear) estimating equations.  

\section{Large-Sample Properties}\label{section-2: large sample properties}

Let $(Y_i,D_i, X_i)$, $i=1, 2, \cdots, n$ be a random sample from $Y\in\mathbb{R}$, $D\in\{0,1\}$ and $ X\in\mathbb{R}^{d_{x}}$. Recall that the probability weight is defined as $e( X)= \Prob[D=1| X]$. Define the conditional moments of the outcome variable as 
\[
\mu_s(e( X))=\Expectation[Y^s|e( X),D=1],\quad s>0,
\]
then the parameter of interest is $\theta_0  = \Expectation[DY/e(X)] = \Expectation[\mu_1(e( X))]$. At this level of generality, we do not attach specific interpretations to the parameter and the random variables in our model. To facilitate understanding, one can think of $Y$ as an observed outcome variable and $D$ as an indicator of treatment status, hence the parameter is the population average of one potential outcome. 

As previewed in Introduction, large-sample properties of the IPW estimator $\hat{\theta}_{n,b_n}$ depend on the tail behavior of the probability weights near zero: if $e(X)$ is bounded away from zero, the IPW estimator is $\sqrt{n}$-consistent and asymptotically Gaussian; in the presence of small probability weights, however, non-Gaussian distributions can emerge. In this section, we first discuss the assumptions and formalize the notion of ``probability weights being close to zero'' or ``having a heavy tail.'' Then we characterize the asymptotic distribution of $\hat{\theta}_{n,b_n}$, and show how it is affected by the trimming threshold $b_n$. 

\subsection{Tail Behavior}\label{subsection-2-1: tail behavior}

For an estimator that takes the form of a sample average (or more generally can be linearized into such), distributional approximation based on the central limit theorem only requires a finite variance. The problem with inverse probability weighting with ``small denominators,'' however, is that the estimator may not have a finite variance. In this case, distributional convergence relies on tail features, which we formalize in the following assumption.

\begin{assumption}[Regular Variation]\label{assumption:tail index}
For some $\gamma_0>1$, the probability weights have a regularly varying tail with index $\gamma_0-1$ at zero:
\begin{align*}
\lim_{t\downarrow 0} \frac{\Prob[e( X)\leq tx]}{\Prob[e( X)\leq t]} = x^{\gamma_0-1},\qquad \text{for all $x>0$}.
\end{align*}
\end{assumption}

Assumption \ref{assumption:tail index} only imposes a local restriction on the tail behavior of the probability weights, and is common when dealing with sums of heavy-tailed random variables. It is equivalent to $\Prob[e( X)\leq x] = c(x)x^{\gamma_0-1}$ with $c(x)$ being a slowly varying function (see the Supplementary Material or \citealt[Chapter XVII]{fellerVol2} for a definition). A special case of Assumption \ref{assumption:tail index} is ``approximately polynomial tail,'' which requires $\lim_{x\downarrow 0}c(x)=c>0$. To see how the tail index $\gamma_0$ features in data, we illustrate in Section \ref{section-4:empirical} with estimated probability weights from an empirical example, and it is clear that the probability weights exhibit a heavy tail near zero. Later in Theorem \ref{thm: IPW asy distribution}, we show that $\gamma_0=2$ is the knife-edge case that separates the Gaussian and the non-Gaussian asymptotic distributions for the (untrimmed) IPW estimator. With $\gamma_0=2$, the probability weights are approximately uniformly distributed, a fact that can be used in practice as a rough guidance on the magnitude of this tail index. 

\begin{remark}[Implied Tail of $ X$]\label{remark:subexponential tail of X}
To see how the tail behavior of the probability weights is related to that of the covariates $ X$, we consider a Logit model: $e( X) = {\exp(X^\Trans \pi_0)}/({1+\exp(X^\Trans \pi_0)})$, which implies $\Prob[e( X)\leq x]= \Prob[ X^\Trans \pi_0 \leq -\log(x^{-1}-1) ]$. As a result, Assumption \ref{assumption:tail index} is equivalent to that, for all $x$ large enough, $\Prob[ X^\Trans \pi_0 \leq -x ]\approx e^{-(\gamma_0-1)x}$, meaning that the (left) tail of $X^\Trans \pi_0$ is approximately sub-exponential. 
\EndOfTheorem
\end{remark}

Assumption \ref{assumption:tail index} characterizes the tail behavior of the probability weights. However, it alone does not suffice for the IPW estimator to have a asymptotic distribution. The reason is that, for sums of random variables without finite variance to converge in distribution, one needs not only a restriction on the shape of the tail, but also a ``tail balance condition.'' For this purpose, we impose the following assumption. 

\begin{assumption}[Conditional Distribution of $Y$]\label{assumption:conditional distribution of true Y}
(i) For some $\epsilon>0$, $\Expectation\big[|Y|^{(\gamma_0\vee2)+\epsilon}\big|e( X)=x, D=1\big]$ is uniformly bounded. (ii) There exists a probability distribution $F$, such that for all bounded and continuous $\ell(\cdot)$, $\Expectation[\ell(Y)|e( X)=x,D=1]\to \int_{\mathbb{R}} \ell(y)F(\diff y)$ as $x\downarrow 0$.  
\end{assumption}

This assumption has two parts. The first part requires the tail of $Y$ to be thinner than that of $D/e( X)$, therefore the tail behavior of $DY/e( X)$ is largely driven by the ``small denominator $e(X)$.'' As our primary focus is the implication of small probability weights entering the IPW estimator rather than a heavy-tailed outcome variable, we maintain this assumption. The second part requires convergence of the conditional distribution of $Y$ given $e( X)$ and $D=1$. Together, they help characterize the tail behavior of $DY/e(X)$. 

\begin{lem}[Tail property of $DY/e( X)$]\label{lemma:tail of DY/E}
Under Assumption \ref{assumption:tail index} and \ref{assumption:conditional distribution of true Y},
\begin{align*}
\lim_{x\to \infty}\frac{x\Prob[DY/e( X) > x]}{\Prob[e( X)<x^{-1}]} &= \frac{\gamma_0-1}{\gamma_0}\alpha_+(0),\quad 
\lim_{x\to \infty}\frac{x\Prob[DY/e( X) < -x]}{\Prob[e( X)<x^{-1}]} = \frac{\gamma_0-1}{\gamma_0}\alpha_-(0),
\end{align*}
where $\alpha_+(x) = \lim_{t\to 0}\Expectation[|Y|^{\gamma_0}\Indicator_{Y> x}\ |e( X)=t, D=1]$ and $\alpha_-(x) = \lim_{t\to 0}\Expectation[|Y|^{\gamma_0}\Indicator_{Y< x}\ |e( X)=t, D=1]$. 
\end{lem}

Assuming the distribution of the outcome variable is nondegenerate conditional on the probability weights being small (i.e., $\alpha_+(0) + \alpha_-(0) > 0$), Lemma \ref{lemma:tail of DY/E} shows that $DY/e(X)$ has regularly varying tails with index $-\gamma_0$. As a result, $\gamma_0$ determines which moment of the IPW estimator is finite: for $s<\gamma_0$, $\Expectation[|DY/e( X)|^s]<\infty$. We compare to a common assumption made in the IPW literature, which requires the probability weights to be bounded away from zero. This assumption is sufficient but not necessary for asymptotic Gaussianity. In fact, the IPW estimator is asymptotically Gaussian as long as $\gamma_0\geq 2$. Intuitively, small denominators appear so infrequently that they will not affect the large-sample properties. For $\gamma_0\in (1,2)$, the IPW estimator no longer has a finite variance, as the distribution of $e(X)$ does not approach zero fast enough (or equivalently, the density of $e(X)$, if it exists, diverges to infinity). This scenario represents the empirical difficulty of dealing with small probability weights entering the IPW estimator, for which regular asymptotic analysis no longer applies. 

Thanks to Assumption \ref{assumption:conditional distribution of true Y}(ii), Lemma \ref{lemma:tail of DY/E} also implies that $DY/e(X)$ has balanced tails: the ratio $\frac{\Prob[DY/e( X) > x]}{\Prob[|DY/e( X)| > x]}$ tends to a finite constant. It turns out that without a finite variance, the asymptotic distribution of the IPW estimator is non-Gaussian, and the asymptotic distribution depends on both the left and right tail of $DY/e(X)$. Thus, tail balancing (and Assumption \ref{assumption:conditional distribution of true Y}(ii)) is indispensable for developing a large-sample theory allowing for small probability weights entering the IPW estimator. 

\subsection{Asymptotic Distribution}\label{subsection-2-2: large sample properties}

The following theorem characterizes the asymptotic distribution of the IPW estimator, both with and without trimming. To make the result concise, we assume the oracle (rather than estimated) probability weights are used, making the IPW estimator a one-step procedure. We extend the theorem to estimated probability weights in the following subsection. In the Supplementary Material, we also discuss how our IPW framework can be generalized to provide robust inference for treatment effect estimands and parameters defined by general (nonlinear) estimating equations.  

\begin{thm}[Asymptotic Distribution]\label{thm: IPW asy distribution}
Assume Assumption \ref{assumption:tail index} and \ref{assumption:conditional distribution of true Y} hold with $\alpha_+(0)+\alpha_-(0)>0$, $b_n\to 0$, and let $a_n$ be defined such that 
\begin{align*}
\frac{n}{a_n^2}\Expectation\Big[ \left|\frac{DY}{e(X)}-\theta_0\right|^2\Indicator_{|DY/e(X)|\leq a_n} \Big]\to 1.
\end{align*}
(i) If $\gamma_0\geq 2$, \eqref{eq:trimmed IPW limiting distribution} holds with $a_{n,b_n}=a_n$, and the asymptotic distribution is standard Gaussian. \\
(ii.1) No trimming, light trimming and moderate trimming: if $\gamma_0< 2$ and $b_na_n\to t\in[0,\infty)$, \eqref{eq:trimmed IPW limiting distribution} holds with $a_{n,b_n}=a_n$, and the asymptotic distribution is infinitely divisible with characteristic function
\begin{align*}
&\ \psi(\zeta) = \exp\left\{ \int_{\mathbb{R}}\frac{e^{i\zeta x} - 1 - i\zeta x}{x^2} M(\diff x) \right\},\\
&\ \qquad \qquad \text{where }M(\diff x) = \diff x \left[\frac{2-\gamma_0}{\alpha_+(0)+\alpha_-(0)}|x|^{1-\gamma_0}\Big(\alpha_+(tx)\Indicator_{x\geq 0} +\alpha_-(tx)\Indicator_{x< 0} \Big)\right].
\end{align*}
(ii.2) Heavy trimming: if $\gamma_0< 2$ and $b_na_n\to \infty$, \eqref{eq:trimmed IPW limiting distribution} holds with $a_{n,b_n} = \sqrt{n\Var[DY/e( X)\Indicator_{e( X)\geq b_n}]}$, and the asymptotic distribution is standard Gaussian. 
\end{thm}

To provide some insight, we first consider the untrimmed IPW estimator ($b_n=0$), whose large-sample properties are summarized in part (i) and (ii.1). Theorem \ref{thm: IPW asy distribution} demonstrates how a non-Gaussian asymptotic distribution can emerge when the untrimmed IPW estimator does not have a finite variance ($\gamma_0<2$). The asymptotic distribution in this case is also known as the L\'{e}vy stable distribution, which has the following equivalent representation,
\begin{align*}
\psi(\zeta) &= -|\zeta|^{\gamma_0} \frac{\Gamma(3-\gamma_0)}{\gamma_0(\gamma_0-1)} \left[-\cos\left(\frac{\gamma_0\pi}{2}\right) + i\frac{\alpha_+(0)-\alpha_-(0)}{\alpha_+(0)+\alpha_-(0)}\mathrm{sgn}(\zeta)\sin\left(\frac{\gamma_0\pi}{2}\right) \right],
\end{align*}
where $\Gamma(\cdot)$ is the gamma function and $\mathrm{sgn}(\cdot)$ is the sign function. From this alternative form, we deduce several properties of the asymptotic L\'{e}vy stable distribution. First, this distribution is not symmetric unless $\alpha_+(0)=\alpha_-(0)$. Second, the characteristic function has a sub-exponential tail, meaning that the limiting L\'{e}vy stable distribution has a smooth density function (although in general it does not have a closed-form expression). Finally, the above characteristic function is continuous in $\gamma_0$, in the sense that as $\gamma_0\uparrow 2$, it reduces to the standard Gaussian characteristic function.

Theorem \ref{thm: IPW asy distribution} also shows how the convergence rate of the untrimmed IPW estimator depends on the tail index $\gamma_0$. For $\gamma_0>2$, the IPW estimator converges at the usual parametric rate $n/a_{n,b_n}\asymp\sqrt{n}$. This extends to the $\gamma_0=2$ case, except that an additional slowly varying factor is present in the convergence rate. For $\gamma_0<2$, the convergence rate is only implicitly defined from a truncated second moment, and generally does not have an explicit formula. One can consider the special case that the probability weights have an approximately polynomial tail: $\Prob[e(X)\leq x]\asymp x^{\gamma_0-1}$, for which $a_{n,b_n}$ can be set to $n^{1/\gamma_0}$. As a result, the untrimmed IPW estimator will have a slower convergence rate if the probability weights have a heavier tail at zero (i.e., smaller $\gamma_0$).  Fortunately, the (unknown) convergence rate is captured by self-normalization (Studentization), which we employ in our robust inference procedure.

Now we discuss the impact of trimming, a strategy commonly employed in practice in response to small probability weights entering the IPW estimator. We distinguish among three cases: light trimming ($b_na_n\to 0$), moderate trimming ($b_na_n\to t\in(0,\infty)$), and heavy trimming ($b_na_n\to \infty$). For light trimming, $b_n$ shrinks to zero fast enough so that asymptotically trimming becomes negligible, and the asymptotic distribution is L\'{e}vy stable as if there were no trimming. For heavy trimming, the trimming threshold shrinks to zero slowly, hence most of the small probability weights are excluded. This leads to a Gaussian asymptotic distribution. The moderate trimming scenario lies between the two extremes. On the one hand, a nontrivial number of small probability weights are discarded, making the limit no longer the L\'{e}vy stable distribution. On the other hand, the trimming is not heavy enough to restore asymptotic Gaussianity. The asymptotic distribution in this case is quite complicated, and depends on two (infinitely dimensional) nuisance parameters, $\alpha_+(\cdot)$ and $\alpha_-(\cdot)$. For this reason, inference is quite challenging. 

Despite the asymptotic distribution taking on a complicated form, moderate trimming as in Theorem \ref{thm: IPW asy distribution}(ii.1) is highly relevant. In Section \ref{subsection-3-1: balancing bias and variance}, we discuss how the trimming threshold can be chosen to balance the bias and variance (i.e., to minimize the mean squared error), which corresponds to this moderate trimming scenario. In addition, unless one employs a very large trimming threshold, it is unclear how well the Gaussian approximation performs in samples of moderate size. 

\subsection{Estimated Probability Weights}\label{subsection-2-3: estimated weights}

The probability weights are usually unknown and are estimated in a first step, which are then plugged into the IPW estimator, making it a two-step estimation problem. Estimating the probability weights in a first step can affect large-sample properties of the IPW estimator through two channels: the estimated weights enter the final estimator both through inverse weighting and through the trimming function. In this subsection, we first impose high-level assumptions and discuss the impact of employing estimated probability weights. Then we verify those high-level assumptions for Logit and Probit models, which are widely used in applied work. 

\begin{assumption}[First-Step Estimation]\label{assumption:first step}
The probability weights are parametrized as $e(X,\pi)$ with $\pi\in \Pi$, and $e(\cdot)$ is continuously differentiable with respect to $\pi$. Let $e(X) = e(X,\pi_0)$ and $\hat{e}(X)=e(X,\hat{\pi}_n)$. Further, (i) $\sqrt{n}(\hat{\pi}_n - \pi_0) = \frac{1}{\sqrt{n}}\sum_{i=1}^n h(D_i,X_i) + \op(1)$, where $h(D_i,X_i)$ is mean zero and has a finite variance; and (ii) For some $\epsilon>0$, $\Expectation\left[\sup_{\pi:|\pi-\pi_0|\leq \epsilon} \left|\frac{e(X_i)}{e(X_i,\pi)^2}\frac{\partial e(X_i,\pi)}{\partial \pi}\right|\right] <\infty$.
\end{assumption}

\begin{assumption}[Trimming Threshold]\label{assumption:trimming threshold}
The trimming threshold satisfies $c_n\sqrt{b_n\Prob[e(X_i)\leq b_n]} \to 0$, where $c_n$ is a positive sequence such that, for any $\epsilon>0$, 
\begin{align*}
c_n^{-1}\max_{1\leq i\leq n}\sup_{|\pi-\pi_0|\leq \epsilon/\sqrt{n}} \left| \frac{1}{e(X_i)}\frac{\partial e(X_i,\pi)}{\partial \pi} \right| = \op(1).
\end{align*} 
\end{assumption} 

Now we state the analogue of Theorem \ref{thm: IPW asy distribution} but with the probability weights estimated in a first step. 

\begin{prop}[Asymptotic Distribution with Estimated Probability Weights]\label{prop:IPW with estimated weights}
Assume Assumption \ref{assumption:tail index}--\ref{assumption:trimming threshold} hold with $\alpha_+(0)+\alpha_-(0)>0$. Let $a_n$ be defined such that
\begin{align*}
\frac{n}{a_n^2}\Expectation\left[ \left|\frac{DY}{e(X)}-\theta_0-A_0h(D,X)\right|^2\Indicator_{|DY/e(X)-A_0h(D,X)|\leq a_n} \right]\to 1,
\end{align*}
where $A_0 = \Expectation\left[ \frac{\mu_1(e(X))}{e(X)} \left.\frac{\partial e(X,\pi)}{\partial \pi}\right|_{\pi=\pi_0} \right]$. Then the IPW estimator has the following linear representation: 
\begin{align*}
\frac{n}{a_{n,b_n}}\left(\hat\theta_{n,b_n} - \theta_0 - \mathsf{B}_{n,b_n}\right) &= \frac{1}{a_{n,b_n}}\sum_{i=1}^n \left(\frac{D_iY_i}{e(X_i)}\Indicator_{e(X_i)\geq b_n} - \theta_0 - \mathsf{B}_{n,b_n} - A_0h(D_i,X_i)\right) + \op(1),
\end{align*}
and the conclusions of Theorem \ref{thm: IPW asy distribution} hold with estimated probability weights. 
\end{prop}

To understand Proposition \ref{prop:IPW with estimated weights}, we consider two cases. In the first case, $\Var[DY/e(X)]<\infty$, and estimating the probability weights in a first step will contribute to the asymptotic variance. The second case corresponds to $\Var[DY/e(X)]=\infty$, implying that the final estimator, $\hat{\theta}_{n,b_n}$, has a slower convergence rate compared to the first-step estimated probability weights. As a result, the two definitions of the scaling factor $a_n$ (in Theorem \ref{thm: IPW asy distribution} and in Proposition \ref{prop:IPW with estimated weights}) are asymptotically equivalent, and the asymptotic distribution will be the same regardless of whether the probability weights are known or estimated. In addition, Proposition \ref{prop:IPW with estimated weights} shows that, despite the estimated probability weights entering both the denominator and the trimming function, the second channel is asymptotically negligible under an additional assumption, which turns out to be very mild in applications.

Assumption \ref{assumption:first step}(i) is standard. In the following remark we provide primitive conditions to justify Assumption \ref{assumption:first step}(ii) and Assumption \ref{assumption:trimming threshold} in Logit and Probit models. 

\begin{remark}[Logit and Probit Models]\label{remark:logit and probit, 1}
Assuming a Logit model for the probability weights, we show in the Supplementary Material that a sufficient condition for Assumption \ref{assumption:first step}(ii) is the covariates having a sub-exponential tail: $\Expectation[e^{\epsilon|X|}]<\infty$ for some (small) $\epsilon>0$. This condition is fully compatible with Assumption \ref{assumption:tail index}, as we show in Remark \ref{remark:subexponential tail of X} that for Assumption \ref{assumption:tail index} to hold in a Logit model, the index $X^\Trans \pi_0$ needs to have a sub-exponential left tail. As for the Probit model, Assumption \ref{assumption:first step}(ii) is implied by a sub-Gaussian tail of the covariates: $\Expectation[e^{\epsilon|X|^2}]<\infty$ for some (small) $\epsilon>0$. Again, it is possible to show that Assumption \ref{assumption:tail index} implies a sub-Gaussian left tail for the index $X^\Trans\pi_0$. 

To verify Assumption \ref{assumption:trimming threshold}, it suffices to set $c_n = \log^2(n)$ for Logit and Probit models. Therefore, we only require the trimming threshold shrinking to zero faster than a logarithmic rate. See the Supplementary Material for details.
\EndOfTheorem
\end{remark}

\section{Robust Inference}\label{section-3: robust inference}

In the previous section, we show that non-Gaussian asymptotic distributions may arise as a result of small probability weights entering the IPW estimator and trimming. In this section, we first study the trimming bias, and show that this bias is usually non-negligible for inference purpose. Together, these findings explain why the point estimate is sensitive to the choice of the trimming threshold, and more importantly, why conventional inference procedures based on the standard Gaussian approximation perform poorly. 

As a remedy, we first introduce a method to choose the trimming threshold by minimizing an empirical mean squared error, and discuss how our trimming threshold selector can be modified in a disciplined way if the researcher prefers to discard more observations. 
Then we propose to combine subsampling with a novel bias correction technique, where the latter employs local polynomial regression to approximate the trimming bias. 

\subsection{Bias, Variance and Trimming Threshold Selection}\label{subsection-3-1: balancing bias and variance}

If the sole purpose of trimming is to stabilize the IPW estimator, one can argue that only a fixed trimming rule, $b_n=b\in(0,1)$, should be used. Such practice, however, completely ignores the bias introduced by trimming, and forces the researcher to change the target estimand and re-interpret the estimation/inference result. Practically, the trimming threshold can be chosen by minimizing the asymptotic mean squared error. We first characterize the bias and variance of the (trimmed) IPW estimator.

\begin{lem}[Bias and Variance of $\hat{\theta}_{n,b_n}$]\label{lem:bias and variance}
Assume Assumption \ref{assumption:tail index} and \ref{assumption:conditional distribution of true Y} hold with $\gamma_0<2$. Further, assume that $\mu_1(\cdot)$ and
$\mu_2(\cdot)$ do not vanish near 0. Then the bias and variance of $\hat\theta_{n,b_n}$ are:
\[\Bias_{n,b_n} = -\mu_1(0)\Prob\left[e( X) \leq b_n\right] (1+o(1)),\qquad \Variance_{n,b_n} 
= \mu_2(0)\frac{1}{n}\Expectation\left[e( X)^{-1}\Indicator_{e( X)\geq b_n}\right](1+o(1)).\]
In addition, $\Bias_{n,b_n}^2/\Variance_{n,b_n}\asymp nb_n\Prob[e( X)\leq b_n]$.
\end{lem} 

To balance the bias and variance, minimizing the leading mean squared error (MSE) with respect to the trimming threshold leads to
\begin{align*}
b_n^\dagger\cdot\Prob[e( X)\leq b_n^\dagger] &= \frac{1}{2n}\frac{\mu_2(0)}{\mu_1(0)^2}.
\end{align*}
The MSE-optimal trimming $b_n^\dagger$ helps understand the three scenarios in Theorem \ref{thm: IPW asy distribution}: light, moderate and heavy trimming. More importantly, it helps clarify whether (and when) the trimming bias features in the asymptotic distribution. (The trimming bias $\Bias_{n,b_n}$ vanishes as long as $b_n\to 0$. What matters for inference, however, is the asymptotic bias, which is defined as the trimming bias scaled by the convergence rate: $\frac{n}{a_{n,b_n}}\mathsf{B}_{n,b_n}$. This asymptotic bias may not be negligible even in large samples.) $b_n^\dagger$ corresponds to the moderate trimming scenario, and since it balances the leading bias and variance, the asymptotic distribution of the trimmed IPW estimator is not centered at the target estimand (i.e., it is asymptotically biased). A trimming threshold that shrinks more slowly than the optimal one corresponds to the heavy trimming scenario, where the bias dominates in the asymptotic distribution. The only scenario in which one can ignore the trimming bias for inference purposes is when light trimming is used. That is, the trimming threshold shrinks faster than $b_n^\dagger$. 

The following theorem shows that, under very mild regularity conditions, the MSE-optimal trimming threshold can be implemented in practice by solving a sample analogue. It also provides a disciplined method for choosing the trimming threshold if the researcher prefers to employ a heavy trimming. 

\begin{thm}[Trimming Threshold Selection]\label{thm:est optimal b}
Assume Assumption \ref{assumption:tail index} holds, and $0<\mu_2(0)/\mu_1(0)^2<\infty$. For any $s>0$, define $b_n$ and $\hat{b}_n$ as:
\begin{align*}
b_n^s\Prob[e( X)\leq b_n]=\frac{1}{2n}\frac{\mu_2(0)}{\mu_1(0)^2},\qquad \hat{b}_n^s\left(\frac{1}{n}\sum_{i=1}^n \Indicator_{e( X)\leq \hat{b}_n}\right)=\frac{1}{2n}\frac{\hat{\mu}_2(0)}{\hat{\mu}_1(0)^2},
\end{align*}
where $\hat{\mu}_1(0)$ and $\hat{\mu}_2(0)$ are some consistent estimates of $\mu_1(0)$ and $\mu_2(0)$, respectively. Then $\hat{b}_n$ is consistent for $b_n$: $\hat{b}_n/b_n\toProb 1$. Therefore, for $0<s<1$, $s=1$ and $s>1$, $\hat{b}_n / b_n^\dagger$ converges in probability to $0$, $1$ and $\infty$, respectively. 

In addition to Assumption \ref{assumption:first step}, if we have for any $\epsilon>0$,
\begin{align*}
\max_{1\leq i\leq n}\sup_{|\pi-\pi_0|\leq \epsilon/\sqrt{n}} \left| \frac{1}{e(X_i)}\frac{\partial e(X_i,\pi)}{\partial \pi} \right| = \op\left(\sqrt{\frac{n}{\log(n)}}\right),
\end{align*} 
then $\hat{b}_n$ can be constructed with estimated probability weights. 
\end{thm}

This theorem states that, as long as one can construct a consistent estimator for the ratio $\mu_2(0)/\mu_1(0)^2$, the optimal trimming threshold can be implemented in practice with the unknown distribution $\Prob[e( X)\leq x]$ replaced by the standard empirical distribution function. In addition, Theorem \ref{thm:est optimal b} allows the use of estimated probability weights to construct $\hat{b}_n$. The extra condition turns out to be quite weak, and is easily satisfied if the probability weights are estimated in a Logit or Probit model. (See Remark \ref{remark:logit and probit, 1}, and the Supplementary Material for further discussion.)

In the following, we show that distributional convergence of the IPW estimator is unaffected by the use of data-driven trimming threshold. Establishing distributional convergence with data-driven tunning parameters tends to be quite difficult in general. In our setting, however, incorporating estimated trimming threshold does not require additional (strong) assumptions, as it is possible to exploit the specific structure that the trimming threshold enters only through an indicator function. 

\begin{prop}[Asymptotic Distribution with Data-Driven Trimming Threshold]\label{prop:data-driven trimming} Assume the assumptions of Theorem \ref{thm:est optimal b} hold. Then Theorem \ref{thm: IPW asy distribution} and Proposition \ref{prop:IPW with estimated weights} hold with data-driven trimming threshold $b_n=\hat{b}_n$.
\end{prop}

\subsection{Bias Correction}\label{subsection-3-2:bias correction}

\begin{figure}[!t]
\centering
\subfloat[]{\resizebox{0.5\columnwidth}{!}{\includegraphics{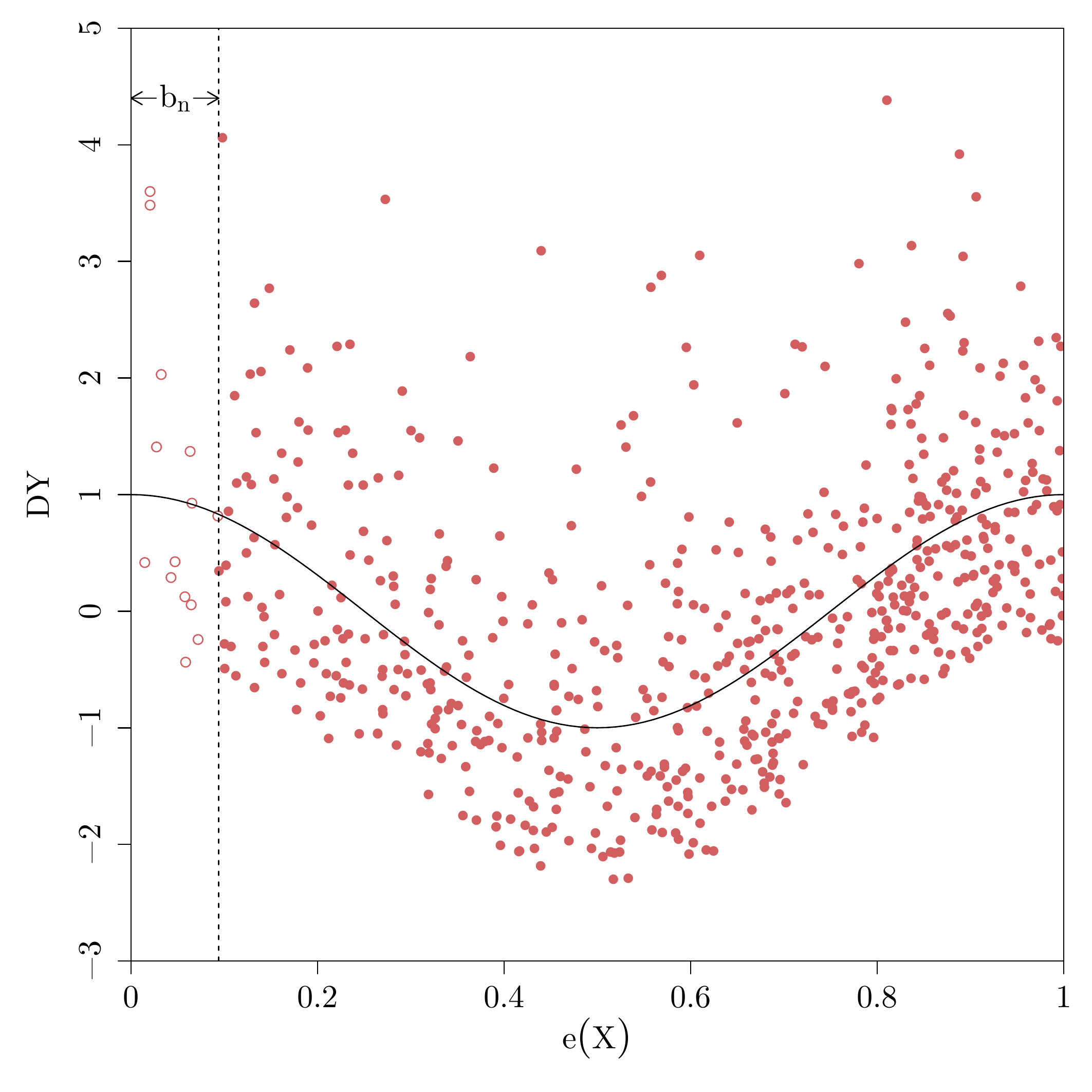}}}
\subfloat[]{\resizebox{0.5\columnwidth}{!}{\includegraphics{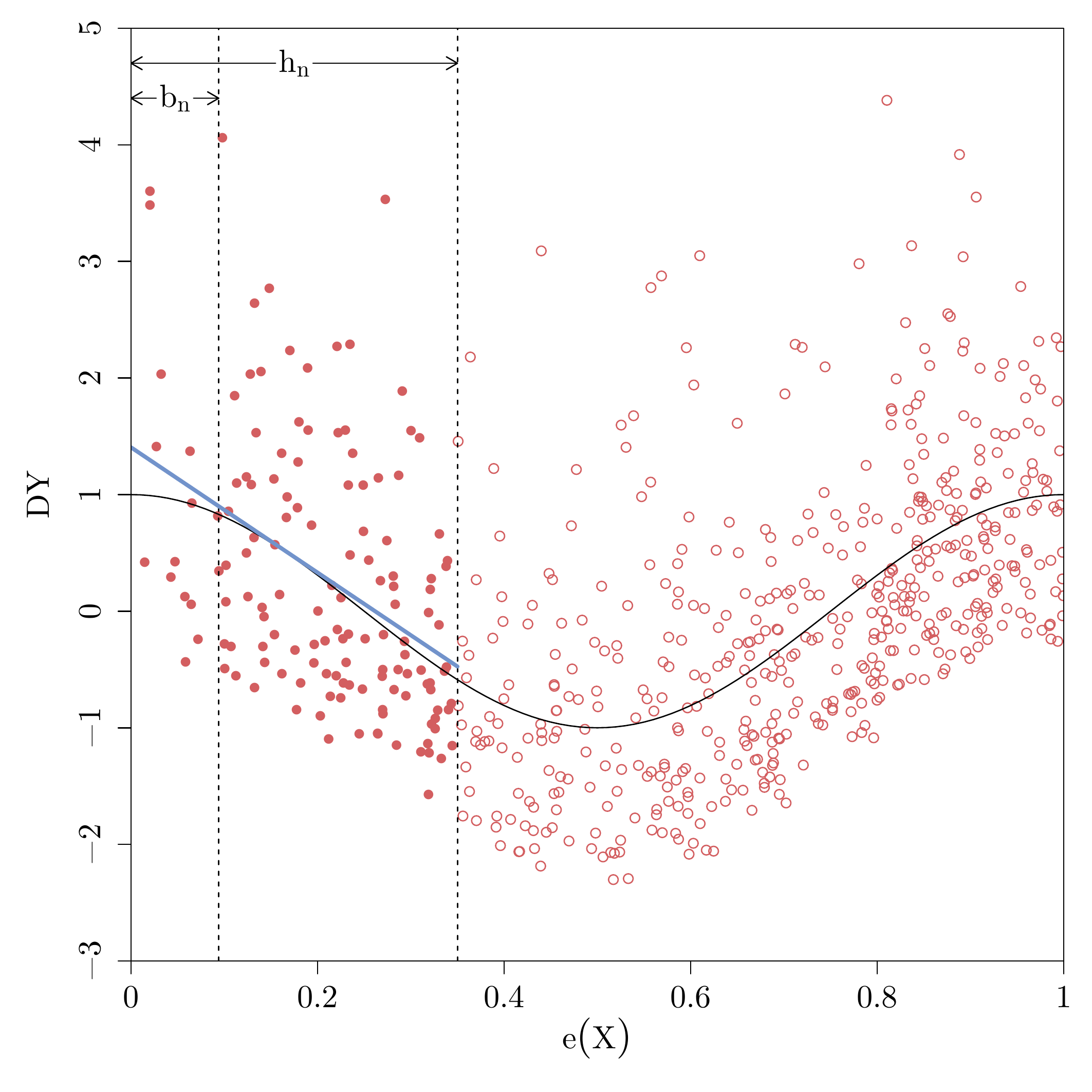}}}
\caption{Trimming and Local Polynomial Bias Estimation. (a) Illustration of Trimming. Circles: trimmed observations. Solid dots: observations included in the estimator. Solid curve: conditional expectation function $\Expectation[Y|e( X),D=1]$. (b) Illustration of the Local Polynomial Regression. Solid dots: observations used in the local polynomial regression. Solid straight line: local linear regression function. }\label{figure:local pol}
\end{figure}

To motivate our bias correction technique, recall that the bias is $\Bias_{n,b_n}=-\Expectation[\mu_1(e( X))\Indicator_{e( X)\leq b_n}]$, where $\mu_1(\cdot)$ is the expectation of the outcome $Y$ conditional on the probability weight and $D=1$. Next, we replace the expectation by a sample average, and the unknown conditional expectation by a $p$-th order polynomial expansion, which is then estimated by local polynomial regressions \citep{fan-Gijbels_1996_Book}. To be precisely, one first implements a $p$-th order local polynomial regression of the outcome variable on the probability weight using the $D=1$ subsample in a region $[0,h_n]$, where $(h_n)_{n\geq 1}$ is a bandwidth sequence. The estimated bias is then constructed by replacing the unknown conditional expectation function and its derivatives by the first-step estimates. Following is the detailed algorithm, which is also illustrated in Figure \ref{figure:local pol}.

\begin{algo}[Bias Estimation]\label{algorithm-1: loc pol bias correction}\ \\
\textbf{Step 1}. With the $D=1$ subsample, regress the outcome variable $Y_i$ on the (estimated) probability weight in a region $[0,h_n]$:
\begin{align*}
\Big[ \hat{\beta}_0, \hat{\beta}_1, \cdots, \hat{\beta}_p \Big]'  = \argmin_{\beta_0,\beta_1,\cdots,\beta_p} \sum_{i=1}^n D_i\Big[Y_i - \sum_{j=0}^p \beta_j \hat{e}( X_i)^j\Big]^2\Indicator_{\hat{e}( X_i)\leq h_n}.
\end{align*}
\textbf{Step 2}. Construct the bias correction term as
\begin{align*}
\hat{\Bias}_{n,b_n} = -\frac{1}{n}\sum_{i=1}^n \left(\sum_{j=0}^p \hat{\beta}_j\hat{e}( X_i)^j\right) \Indicator_{\hat{e}( X_i)\leq b_n},
\end{align*}
so that the bias-corrected estimator is $\hat{\theta}_{n,b_n}^{\mathsf{bc}} = \hat{\theta}_{n,b_n} - \hat{\Bias}_{n,b_n}$.
\EndOfTheorem
\end{algo}

By inspecting the bias-corrected estimator, our procedure can be understood as a ``local regression adjustment,'' since we replace the trimmed observations by its conditional expectation, which is further approximated by a local polynomial. In the local polynomial regression step, it is possible to incorporate other kernel functions: we use the uniform kernel to avoid introducing additional notation, but all the main conclusions continue to hold with other commonly employed kernel functions. As for the order of local polynomial regression, common choices are $p=1$ and $2$, which reduce the bias to a satisfactory level without introducing too much additional variation. 

Standard results form the local polynomial regression literature require the density of the design variable to be bounded away from zero, which is not satisfied in our context. In the $D=1$ subsample which we use for the local polynomial regression, the distribution of the probability weights quickly vanishes near the origin. (More precisely, $\Prob[e(X)\leq x|D=1]\prec x$ as $x\downarrow 0$, meaning that the conditional density of the probability weights (if it exists) tends to zero: $f_{e(X)|D=1}(0)=0$.) As a result, nonstandard scaling is needed to derive large-sample properties of $\hat{\mu}_1^{(j)}(0)$. See the Supplementary Material for a precise statement.  

\begin{thm}[Large-Sample Properties of the Estimated Bias]\label{thm:bias correction validity}
Assume Assumption \ref{assumption:tail index} and \ref{assumption:conditional distribution of true Y} (and in addition Assumption \ref{assumption:first step} and \ref{assumption:trimming threshold} with estimated probability weights) hold. Further, assume (i) $\mu_1(\cdot)$ is $p+1$ times continuously differentiable; (ii) $\mu_2(0)-\mu_1(0)^2>0$; (iii) the bandwidth sequence satisfies $nh_n^{2p+3}\Prob[e( X)\leq h_n] \asymp 1$; (iv) $nb_n^{2p+3}\Prob[e( X)\leq b_n]\to 0$. Then the bias correction is valid, and does not affect the asymptotic distribution: $\hat{\theta}_{n,b_n}^{\mathsf{bc}} - \theta_0 = (\hat{\theta}_{n,b_n} - \Bias_{n,b_n} - \theta_0)(1+\op(1))$.
\end{thm}

Theorem \ref{thm:bias correction validity} has several important implications. First, our bias correction is valid for a wide range of trimming threshold choices, as long as the trimming threshold does not shrink to zero too slowly: $nb_n^{2p+3}\Prob[e( X)\leq b_n]\to 0$. However, fixed trimming $b_n=b\in(0,1)$ is ruled out. This is not surprising, since under fixed trimming the correct scaling is $\sqrt{n}$, and generally the bias cannot be estimated at this rate without additional parametric assumptions. 

Second, it gives a guidance on how the bandwidth for the local polynomial regression can be chosen. In practice, this is done by solving $n\hat{h}_n^{2p+3}\hat{\Prob}[e( X)\leq \hat{h}_n]=c$ for some $c>0$, so that the resulting bandwidth makes the (squared) bias and variance of the local polynomial regression the same order. A simple strategy is to set $c=1$. It is also possible to construct a bandwidth that minimizes the leading mean squared error of the local polynomial regression, for which $c$ has to be estimated in a pilot step (see the Supplementary Material for a characterization of the leading bias and variance). 

Third, it shows how trimming and bias correction together can help improve the convergence rate of the (untrimmed) IPW estimator. From Theorem \ref{thm: IPW asy distribution}(ii), we have $|\hat{\theta}_{n,b_n}-\theta_0-\Bias_{n,b_n}|=\Op((n/a_{n,b_n})^{-1})$, where the convergence rate $n/a_{n,b_n}$ is typically faster when a heavier trimming is employed. This, however, should not be interpreted as a real improvement, as the trimming bias can be quite large. With bias correction, it is possible to achieve a faster rate of convergence for the target estimand, since under the assumptions of Theorem \ref{thm:bias correction validity}, one has $|\hat{\theta}_{n,b_n}^{\mathsf{bc}}-\theta_0|=\Op((n/a_{n,b_n})^{-1})$, which is valid for a wide rage of trimming threshold choices.  

Finally, we note that when Assumption \ref{assumption:tail index} is violated, bias correction may not be feasible. For example, if in some region of the covariate distribution there are lots of observations from one group but not the other, there will be a spike very close zero (or at zero) in the probability weights distribution. As a result, bias correction in either case requires extrapolating a local polynomial regression, which can be unreliable. 

\subsection{Robust Inference}\label{subsection-3-3: inference}

The asymptotic distribution of the IPW estimator can be quite complicated and depend on multiple nuisance parameters which are usually difficult to estimate. We propose the use of subsampling, which is a powerful data-driven method for distributional approximation. It draws samples of size $m\ll n$ and recomputes the statistic with each subsample. Together with our bias correction technique, subsampling can be employed to conduct statistical inference and to construct confidence intervals that are valid for the target estimand. Although Theorem \ref{thm:bias correction validity} states that estimating the bias does not have a first order contribution to the asymptotic distribution, it may still introduce additional variability in finite samples \citep*{calonico2018effect}. Therefore, we recommend subsampling the bias-corrected statistic. 

\begin{algo}[Robust Inference]\label{algorithm-2: robust inference}
Let $\hat{\theta}_{n,b_n}^{\mathsf{bc}}$ be defined as in Algorithm \ref{algorithm-1: loc pol bias correction}, and 
\begin{align*}
T_{n,b_n}=\frac{\hat{\theta}_{n,b_n}^{\mathsf{bc}}-\theta_0}{S_{n,b_n}/\sqrt{n}},\qquad
S_{n,b_n} = \sqrt{\frac{1}{n-1}\sum_{i=1}^n \left( \frac{D_iY_i}{\hat{e}( X_i)}\Indicator_{\hat{e}( X_i)\geq b_n} - \hat{\theta}_{n,b_n} \right)^2}.
\end{align*}
\textbf{Step 1}. Draw $m\ll n$ observations from the original data without replacement, denoted by $(Y_i^\star,D_i^\star,X_i^\star)$, $i=1,2,\cdots, m$. \\
\textbf{Step 2}. Construct the trimmed IPW estimator and the bias correction term from the new subsample, and write the bias-corrected and self-normalized statistic as
\begin{align*}
T_{m,b_m}^\star &= \frac{\hat{\theta}_{m,b_m}^{\star\mathsf{bc}} - \hat{\theta}_{n,b_n}^{\mathsf{bc}} }{S_{m,b_m}^\star/\sqrt{m}},\qquad S_{m,b_m}^\star = \sqrt{\frac{1}{m-1}\sum_{i=1}^m \left( \frac{D_i^\star Y_i^\star}{\hat{e}^\star( X_i^\star)}\Indicator_{\hat{e}^\star( X_i^\star)\geq b_m} - \hat{\theta}_{m,b_m}^\star \right)^2}.
\end{align*}
\textbf{Step 3}. Repeat Step 1 and 2, and a $(1-\alpha)\%$-confidence interval can be constructed as
\begin{align*}
\left[ \hat{\theta}_{n,b_n}^{\mathsf{bc}} - q_{1-\frac{\alpha}{2}}(T_{m,b_m}^\star)\frac{S_{n,b_n}}{\sqrt{n}}\quad,\quad \hat{\theta}_{n,b_n}^{\mathsf{bc}} - q_{\frac{\alpha}{2}}(T_{m,b_m}^\star)\frac{S_{n,b_n}}{\sqrt{n}}  \right],
\end{align*}
where $q_{(\cdot)}(T^\star_{m,b_m})$ denotes the quantile of the statistic $T^\star_{m,b_m}$.
\EndOfTheorem
\end{algo}

Subsampling validity typically relies on the existence of an asymptotic distribution \citep{politis1994large,romano1999subsampling}. We follow this approach and justify our robust inference procedure by showing that the self-normalized statistic $T_{n,b_n}$ converges in distribution. Under $\gamma_0>2$, $S_n$ converges in probability and $T_{n,b_n}$ converges to a Gaussian distribution. Asymptotic Gaussianity of $T_{n,b_n}$ continues to hold for $\gamma_0=2$. Under $\gamma_0<2$, $T_{n,b_n}$ still converges in distribution, although the limit will depend on the trimming threshold. Establishing the asymptotic distribution in the heavy trimming case is relatively easy (Lindeberg-Feller central limit theorem). With light or moderate trimming, however, the asymptotic distribution of $T_{n,b_n}$ is quite complicated. This technical by-product generalizes \cite*{logan1973limit}. (To be precise, with light trimming, we obtain the same distribution as in \cite{logan1973limit}, while the asymptotic distribution of $T_{n,b_n}$ with moderate trimming is new.) We leave the details to the Supplementary Material.

\begin{thm}[Validity of Robust Inference]\label{thm:subsampling validity}
Assume the assumptions of Theorem \ref{thm: IPW asy distribution} (or Proposition \ref{prop:IPW with estimated weights} with estimated probability weights) and Theorem \ref{thm:bias correction validity} hold, $m\to \infty$, and $m/n\to 0$. Then $\sup_{t\in\mathbb{R}}| \Prob[T_{n,b_n}\leq t] - \Prob^\star[T_{m,b_m}^\star \leq t] | \toProb 0$. 
\end{thm}

Before closing this section, we address two practical issues when applying the robust inference procedure. First, it is desirable to have an automatic and adaptive procedure to capture the possibly unknown convergence rate $n/a_{n,b_n}$, as the convergence rate depends on the tail index $\gamma_0$. In the subsampling algorithm, this is achieved by self-normalization (Studentization). Second, one has to choose the subsample size $m$. Some suggestions have been made in the literature: \cite{arcones1989bootstrap} suggest to use $m=\lfloor n/\log\log(n)^{1+\epsilon}\rfloor$ for some $\epsilon>0$, although they consider the $m$-out-of-$n$ bootstrap. \cite{romano1999subsampling} propose a calibration technique. We use $m=\lfloor n/\log (n)\rfloor$ for our simulation study in the Supplementary Material, which performs reasonably well. 

\section{Empirical Illustration}\label{section-4:empirical}
 
In this section, we revisit a dataset from the National Supported Work (NSW) program. Our aim is not to discuss to what extent experimental estimates can be recovered by non-experimental methods. Rather, we use this dataset to showcase our robust inference procedure. The NSW program is a labor training program implemented in 1970's by providing work experience to selected individuals. It has been analyzed in multiple studies since \cite{lalonde1986evaluating}. We use the same dataset employed in \cite{dehejia1998rcausal}. Our sample consists of the treated individuals in the NSW experimental group ($D=1$, sample size $n_1=185$), and a nonexperimental comparison group from the Panel Study of Income Dynamics (PSID, $D=0$, sample size $n_0=1,157$). The outcome variable $Y$ is the post-intervention earning measured in 1978. The covariates $X$ include information on age, education, marital status, ethnicity and earnings in 1974 and 1975. We refer interested readers to \cite{dehejia1998rcausal,dehejia2002propensity} and \cite{smith2005does} for more details on variable definition and sample inclusion. We follow the literature and focus on the treatment effect on the treated (ATT), 
\begin{align*}
\hat{\tau}_{n,b_n}^{\mathtt{ATT}} &= \frac{1}{n_1}\sum_{i=1}^n \left[D_iY_i - \frac{\hat{e}(X_i)}{1-\hat{e}(X_i)}(1-D_i)Y_i\Indicator_{1-\hat{e}(X_i)\geq b_n}\right],
\end{align*}
which requires weighting observations from the comparison group by $\hat{e}(X)/(1-\hat{e}(X))$. As a result, probability weights that are close to 1 can pose a challenge to both estimation and inference. (We discuss in the Supplemental Material how our IPW framework can be generalized to provide robust inference for treatment effect estimands.)

\begin{figure}[!t]
\centering
\resizebox{0.49\columnwidth}{!}{\includegraphics{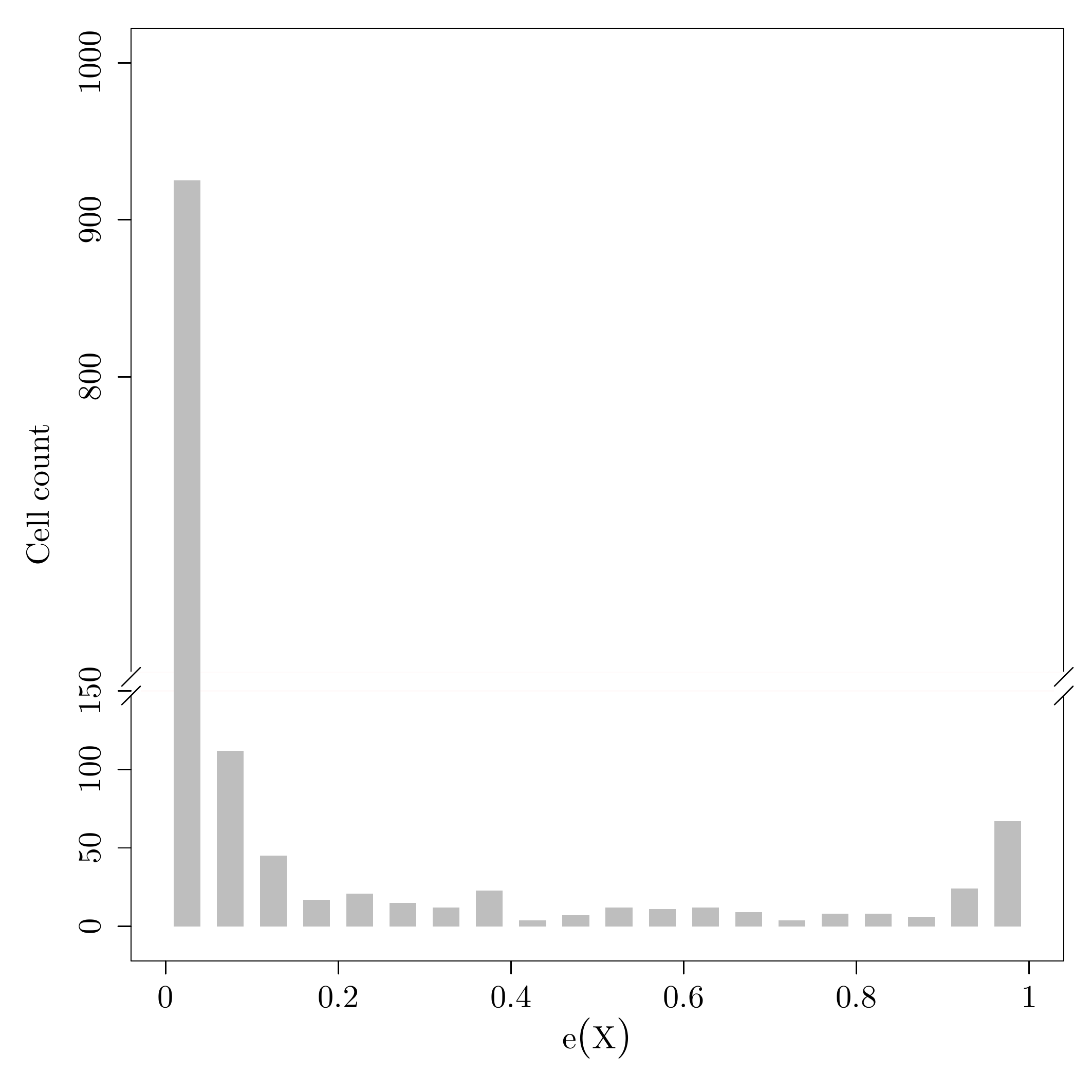}}
\resizebox{0.49\columnwidth}{!}{\includegraphics{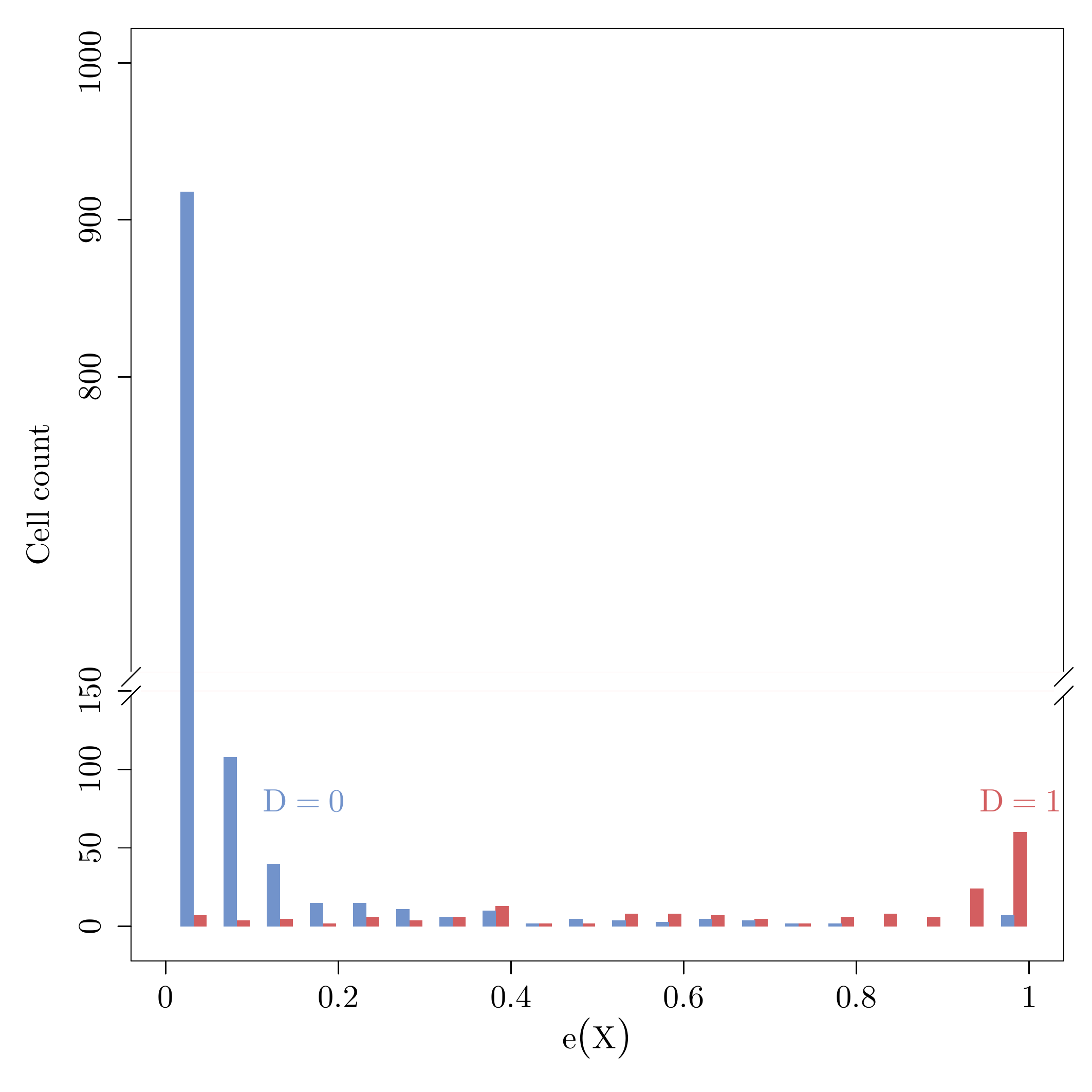}}
\caption{Histogram of the Estimated Probability Weights.}\label{subfig:estimated p-score}
\end{figure}

The probability weight is estimated in a Logit model with $\mathtt{age}$, $\mathtt{education}$, $\mathtt{earn1974}$, $\mathtt{earn1975}$, $\mathtt{age}^2$, $\mathtt{education}^2$, $\mathtt{earn1974}^2$, $\mathtt{earn1975}^2$, three indicators for $\mathtt{married}$, $\mathtt{black}$ and $\mathtt{hispanic}$, and an interaction term $\mathtt{black}\times \mathtt{u74}$, where $\mathtt{u74}$ is the unemployment status in 1974. Figure \ref{subfig:estimated p-score} plots the distribution of the estimated probability weights, which clearly exhibits a heavy tail near 1. Since $\gamma_0=2$ roughly corresponds to uniformly distributed probability weights, the tail index in this dataset should be well below $2$, suggesting that standard inference procedures based on the Gaussian approximation may not perform well. 

\begin{figure}[!t]
\centering
\resizebox{1\columnwidth}{!}{\includegraphics{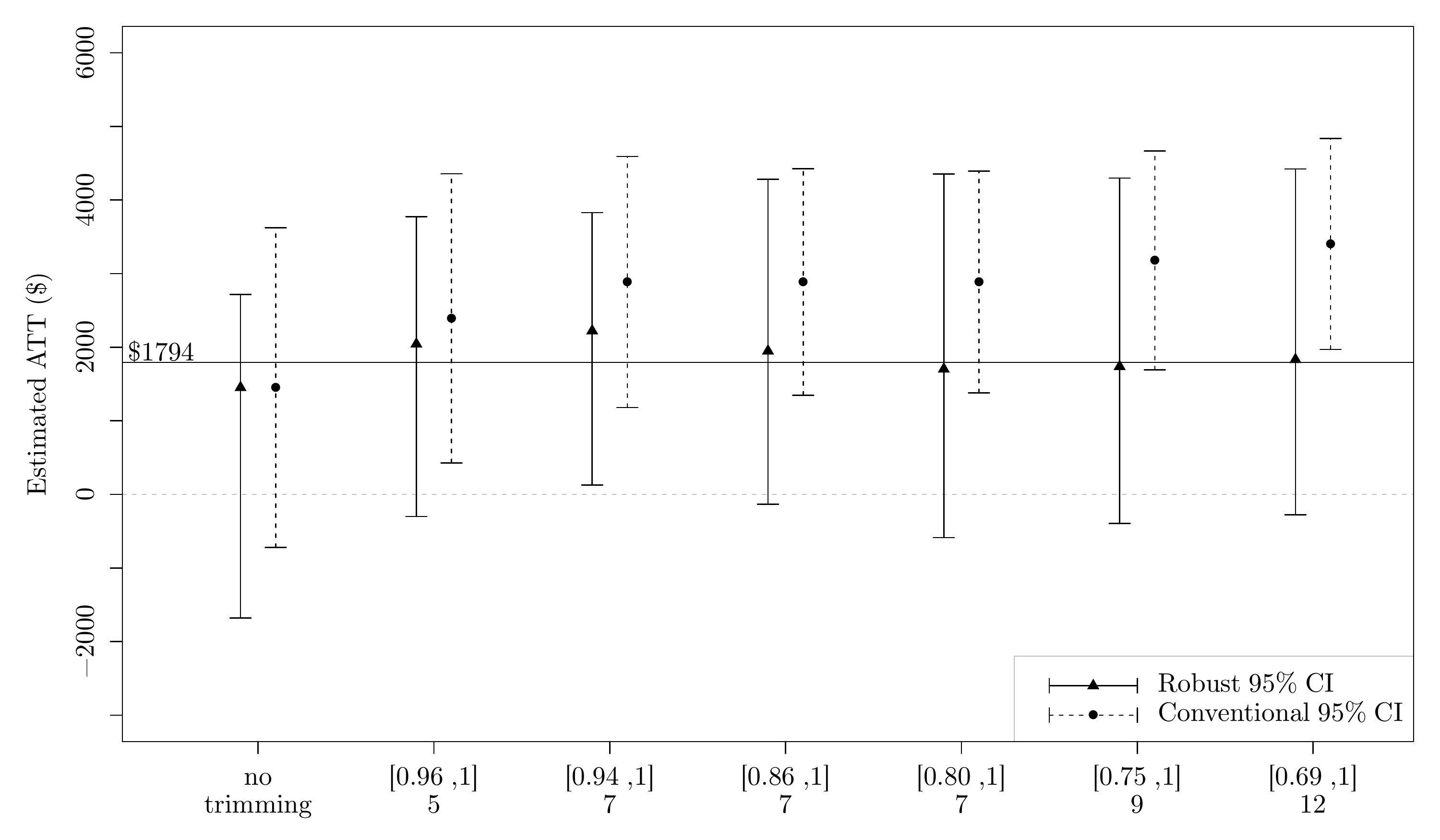}}
\caption{Treatment Effect on the Treated Estimation and Inference. Numbers below the horizontal axis show the trimming threshold/region and the effective number of observations trimmed from the comparison group. The experimental benchmark ($\$1,794$) is indicated by the solid horizontal line. }\label{figure:empirical}
\end{figure}

In Figure \ref{figure:empirical}, we plot the bias-corrected ATT estimates (solid triangles) and the robust 95\% confidence intervals (solid vertical lines) with different trimming thresholds. For comparison, we also show conventional point estimates and confidence intervals (solid dots and dashed vertical lines, based on the Gaussian approximation) using the same trimming thresholds. Without trimming, the point estimate is $\$1,451$ with a confidence interval $[-1,763,\ 2,739]$. The robust confidence interval is asymmetric around the point estimate, which is a feature also predicted by our theory. For the trimmed IPW estimator, the trimming thresholds are chosen following Theorem \ref{thm:est optimal b}, and the region used for local polynomial bias estimation is $[0.71, 1]$, corresponding to a bandwidth $h_n=0.29$. Under the mean squared error optimal trimming, units in the comparison group with probability weights above $0.96$ (five observations) are discarded. Compared to the untrimmed case, the robust confidence interval becomes more symmetric. 

In this empirical example, a noteworthy feature of our method is that both the bias-corrected point estimates and the robust confidence intervals remain quite stable for a range of trimming threshold choices, and the point estimates are very close to the experimental benchmark ($\$1,794$). This is in stark contrast to conventional confidence intervals that rely on Gaussian approximation. First, conventional confidence intervals fail to adapt to the non-Gaussian asymptotic distributions we documented in Theorem \ref{thm: IPW asy distribution}, and are overly optimistic/narrow. Second, by ignoring the trimming bias, they are only valid for a pseudo-true parameter implicitly defined by the trimming threshold. As a result, the researcher changes the target estimand each time a different trimming threshold is used, making conventional confidence intervals very sensitive to $b_n$. 

\section{Conclusion}\label{section:conclusion}

We study the large-sample properties of the Inverse Probability Weighting (IPW) estimator. We show that, in the presence of small probability weights, this estimator may have a slower-than-$\sqrt{n}$ convergence rate and a non-Gaussian asymptotic distribution. We also study the effect of discarding observations with small probability weights, and show that such trimming not only complicates the asymptotic distribution, but also causes a non-negligible bias. We propose an inference procedure that is robust not only to small probability weights entering the IPW estimator but also to a range of trimming threshold choices. The ``two-way robustness'' is achieved by combining resampling with a novel local polynomial-based bias-correction technique. We also propose a method to choose the trimming threshold, and show that our inference procedure remains valid with the use of a data-driven trimming threshold.

To conserve space, we report additional results and simulation evidence in the online Supplementary Material. In particular, we discuss there how our IPW framework can be generalized to provide robust inference for treatment effect estimands and parameters defined by general (nonlinear) estimating equations.

\singlespacing
\bibliographystyle{jasa}
\bibliography{Ma-Wang-2019-RobustIPW--References}

\clearpage

\end{document}